\documentclass[prb,aps,superscriptaddress,floatfix,tightenlines,showpacs,twocolumn]{revtex4-2}
\usepackage{graphicx}
\usepackage{dcolumn}   
\usepackage{bm}        
\usepackage{amssymb}   
\usepackage{amsmath}
\usepackage{url}
\usepackage{layout}
\usepackage{color}
\usepackage{colortbl}
\usepackage{epsfig}

\usepackage{mathrsfs}

\usepackage{appendix}
\usepackage{xcolor}
\usepackage[thinlines]{easytable}
\usepackage{makecell}
\usepackage{tikz,pgf}
\usepackage{booktabs}
\usepackage{float}
\usepackage{hyperref}
\hypersetup{colorlinks,citecolor=blue}

\begin{document}
\title{Spin-density wave and superconductivity in La$_4$Ni$_3$O$_{10}$ under ambient pressure}

\author{Ming Zhang}
\thanks{These three authors contributed equally to this work.}
\affiliation{Zhejiang Key Laboratory of Quantum State Control and Optical Field Manipulation,
Department of Physics, Zhejiang Sci-Tech University, 310018 Hangzhou, China}

\author{Hongyi Sun}
\thanks{These three authors contributed equally to this work.}
\affiliation{Shenzhen Institute for Quantum Science and Engineering, Southern University of Science and Technology, Shenzhen 518055, China}
\affiliation{International Quantum Academy, Shenzhen 518048, China}

\author{Yu-Bo Liu}
\thanks{These three authors contributed equally to this work.}
\affiliation{School of Physics, Beijing Institute of Technology, Beijing 100081, China}

\author{Qihang Liu}
\affiliation{Department of Physics and Guangdong Basic Research Center of Excellence for Quantum Science, Southern University of Science and Technology, Shenzhen 518055, China}
\affiliation{Shenzhen Key Laboratory of Advanced Quantum Functional Materials and Devices, Southern University of Science and Technology, Shenzhen 518055, China}

\author{Wei-Qiang Chen}
\email{chenwq@sustech.edu.cn}
\affiliation{Department of Physics and Guangdong Basic Research Center of Excellence for Quantum Science, Southern University of Science and Technology, Shenzhen 518055, China}
\affiliation{Shenzhen Key Laboratory of Advanced Quantum Functional Materials and Devices, Southern University of Science and Technology, Shenzhen 518055, China}

\author{Fan Yang}
\email{yangfan\_blg@bit.edu.cn}
\affiliation{School of Physics, Beijing Institute of Technology, Beijing 100081, China}

\begin{abstract}
We investigate the spin-density wave (SDW) behavior and the potential for superconductivity (SC) in La$_4$Ni$_3$O$_{10}$ under ambient pressure using a multi-orbital random-phase approximation (RPA). Starting with a twelve-orbital tight-binding model derived from density functional theory (DFT) calculations, we explore the influence of Hubbard interactions on SDW formation. Our analysis reveals a stripe-like SDW characterized by an incommensurate wave vector, $\mathbf{Q}\approx(\pm 0.7\pi,0)$, suggesting a possible density wave instability in agreement with recent experiments. This configuration is driven by nesting of outer-layer Ni $d_{z^2}$ orbitals and exhibits interlayer antiferromagnetic ordering between the top and bottom NiO layers, with the middle layer serving as a node. We demonstrate that the Hund's coupling $J_H$ is the primary driver of the observed SDW. While superconductivity is absent in the undoped system under ambient pressure, it becomes attainable with appropriate hole doping ($\delta=-0.4$), resulting in a SC gap structure similar to the high-pressure phase. Our study identifies the specific conditions for realizing the ambient pressure stripe density wave: $J_H>0.16U$. Additionally, when doping leads to sufficient nesting at (0,0), the system’s magnetic fluctuations transition into  a stable Néel-type antiferromagnetic state, analogous to the high-pressure case.
\end{abstract}
\maketitle

\section{Introduction}
The discovery of superconductivity at 80 K in La$_3$Ni$_2$O$_{7}$ under pressure has solidified the nickelates' place within the high-temperature superconductor family\cite{Wang2023LNO,YuanHQ2023LNO,Wang2023LNOb,wang2023LNOpoly,wang2023la2prnio7,zhang2023pressure}. More recently, La$_4$Ni$_3$O$_{10}$, another Ruddlesden-Popper (RP) phase compound, has been found to exhibit superconductivity under pressure, with a transition temperature ($T_c$) of approximately 25-30 K\cite{zhu2024superconductivity,zhang2023superconductivity,huang2024signature,li2023trilayer}. While La$_4$Ni$_3$O$_{10}$ has a lower $T_c$ than La$_3$Ni$_2$O$_{7}$, it exhibits a much higher superconducting volume fraction\cite{zhu2024superconductivity}, suggesting bulk superconductivity, in contrast to the filamentary behavior observed in La$_3$Ni$_2$O$_{7}$\cite{zhou2023evidence}. Along with the earlier discovery of superconductivity in infinite-layer nickelates (Nd$_{1-x}$Sr$_x$NiO$_2$, $T_c\sim$9-15 K)\cite{li2019superconductivity,lee2023linear,nomura2022superconductivity,gu2022superconductivity}, these findings have sparked interest in the electron correlations and pairing mechanisms in nickelates\cite{yang2023arpes,wang2023structure,cui2023strain,sui2023rno,chen2024electronic,YaoDX2023,Dagotto2023,WangQH2023,lechermann2023,Kuroki2023,HuJP2023,ZhangGM2023DMRG,Werner2023,shilenko2023correlated,WuWei2023charge,cao2023flat,chen2023critical,YangF2023,lu2023bilayertJ,oh2023type2,zhang2023structural,liao2023electron,qu2023bilayer,Yi_Feng2023,jiang2023high,zhang2023trends,huang2023impurity,qin2023high,tian2023correlation,jiang2023pressure,lu2023sc,kitamine2023,luo2023high,zhang2023strong,pan2023rno,sakakibara2023La4Ni3O10,lange2023mixedtj,geisler2023structural,yang2023strong,rhodes2023structural,lange2023feshbach,kaneko2023pair,zhang2023la3ni2o6,ouyang2023hund,heier2023competing,yuan2023trilayer,li2024structural,fan2023sc,geisler2024optical,wu2024deconfined,wang2024electronic,botzel2024theory,Yang2024effective,Li2024ele,PhysRevB.110.L060506,wang2024bulk,li2024distinguishing,jiang2024theory,zhou2024revealing,wang2024chemical,chen2024tri,yang2024decom,ryee2024quenched,Chen2024poly,Dong2024vis,Lu2024interplay,qin2024frustrated,seo1996electronic}. Despite the limited number of known nickelate superconductors, their shared RP structure and $T_c$ values comparable to those of cuprates suggest the potential of nickelates as a promising area for further exploration in high-temperature superconductivity.

Similar to cuprates and iron-based superconductors\cite{ichikawa2000local,dai2015antiferromagnetic}, La$_3$Ni$_2$O$_{7}$ and La$_4$Ni$_3$O$_{10}$ exhibit complex competing orders, including density wave transitions and superconductivity\cite{XIE20243221,du2024correlated,Kakoi2024,xie2024neutron,labollita2023ele,Fukamachi2001,feng2024unaltered,meng2024density,fan2024tunn,chen2024non,lin2024magnetic,xu2024pressure,LI2024distinct,Leonov2024Electronic}. Understanding the density wave instability in these compounds is essential for elucidating the mechanism of superconductivity. At ambient pressure, both La$_3$Ni$_2$O$_{7}$ and La$_4$Ni$_3$O$_{10}$ display density wave transitions. In La$_3$Ni$_2$O$_{7}$, $\mu$SR\cite{khasanov2024pressure,chen2024evidence}, NMR\cite{dan2024spin}, and RIXS\cite{chen2024electronic} measurements reveal a spin density wave (SDW) transition around 150 K, while optical conductivity data indicate the emergence of an energy gap below 115 K, suggesting a possible charge-related density wave transition\cite{Wang2022LNO,liu2024electronic,wu2001magnetic,gupta2024anisotropic}. In La$_4$Ni$_3$O$_{10}$, X-ray and neutron scattering studies at ambient pressure have shown the emergence of an intertwined SDW and charge density wave (CDW) on the Ni sublattice below 135 K, accompanied by an unusual metal-to-metal transition\cite{zhang2020intertwined,xu2024origin}. These experimental findings on density waves in La$_3$Ni$_2$O$_{7}$ and La$_4$Ni$_3$O$_{10}$ have also sparked several theoretical studies and discussions\cite{zhang2024emergent,Leonov2024Electronic,labollita2024assessing,qin2024intertwined,wang2024electronic,Yi2024nature,ni2024first,huo2024electronic,Leonov20244310,tian2024effective,wang2024non}.

For La$_4$Ni$_3$O$_{10}$ at ambient pressure, current theoretical studies suggest that it exhibits complex magnetic fluctuation characteristics. The outer Ni layers show stronger spin fluctuations, which may contribute to "strange metal" behavior, while the inner Ni layers display weaker spin fluctuations. These phenomena are driven by layer-dependent Hund’s rule coupling and competition with crystal field splitting, influenced by electronic correlations and the hybridization of the Ni $d_{z^2}$ and $d_{x^2-y^2}$ orbitals\cite{huo2024electronic,Leonov20244310,tian2024effective,wang2024non}. However, there is a lack of theoretical investigation into the non-commensurate wave vectors observed experimentally\cite{zhang2020intertwined} and the specific effects of Hund’s rule coupling on the magnetic fluctuation characteristics.

In this study, we investigate the SDW behavior and its potential for superconductivity in ambient pressure La$_4$Ni$_3$O$_{10}$ using the random-phase approximation (RPA) approach. Starting with a twelve-orbital tight-binding (TB) model derived from density functional theory (DFT)  calculations, we incorporate multi-orbital Hubbard interactions to explore the system's properties. Our RPA analysis reveals a stripe SDW characterized by a wave vector $\mathbf{Q}\approx(\pm 0.7\pi,0)$, corresponding to the nesting vector between the $\alpha_1$-pocket and $\beta_1$-pockets. Both pockets are dominated by contributions from the outer-layer Ni $d_{z^2}$ orbitals. In the trilayer structure, the top and bottom layers exhibit interlayer antiferromagnetic ordering, with the middle layer acting as an SDW node. Notably, the magnetic moments of the $d_{z^2}$ orbitals are significantly larger than those of the $d_{x^2-y^2}$ orbitals. Furthermore, we find that achieving this non-commensurate SDW wave vector requires a Hund’s rule coupling $J_H>0.16U$. Upon considering appropriate hole doping, the system is predicted to transition into a superconducting phase.

The rest of this paper is organized as follows: In Sec. \ref{sec:TB}, we construct an effective TB model based on the DFT band structure. Sec. \ref{sec:SDW} presents our calculation of the SDW in ambient-pressure La$_4$Ni$_3$O$_{10}$ using a multi-orbital RPA method, detailing both inter-unit-cell and intra-unit-cell pattern. In Sec. \ref{sec:CompHP}, we compared the Fermi surface (FS) nesting and density of states (DOS) features at ambient and high pressures, analyzing the mechanisms underlying the formation of the stripe-type SDW at ambient pressure and the Néel-type SDW at high pressure. Sec. \ref{Doping} demonstrates that appropriate hole doping can induce superconductivity at ambient pressure. A summary is given in Sec. \ref{sec:conclusion} together with some discussions about possible experimental implications.

\section{Band Structure and TB model}
\label{sec:TB}
\begin{figure}[htbp]
\centering
\includegraphics[width=0.45\textwidth]{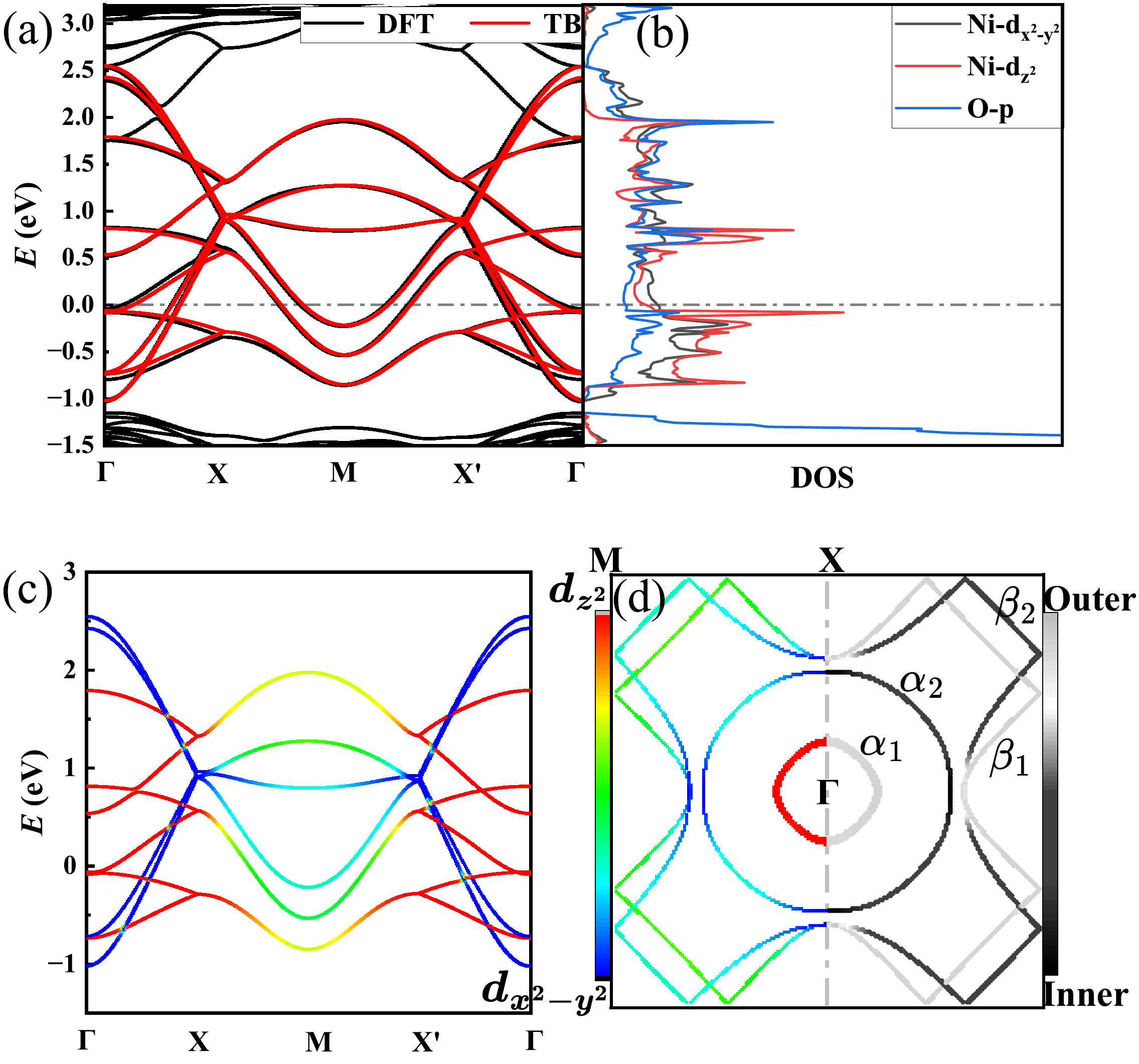}
\caption{(color online) Band structure of the DFT and twelve-orbital TB model for ambient pressure La$_4$Ni$_3$O$_{10}$, using experimentally refined lattice constants. (a) Comparison of the DFT band structure (black solid) and the twelve-orbital TB model (red solid) along the high-symmetry lines of $\text{La}_4\text{Ni}_3\text{O}_{10}$. (b) The DOS for various orbital components from the DFT band in (a). (c) The TB band structure corresponding to (a). (d) FS in the Brillouin zone (BZ), with the four pockets labeled. The color scheme in (c) and the left half of (d) represents the relative contributions of the $d_{z^2}$ and $d_{x^2-y^2}$ orbital, while the colors scheme on the right half of (d) indicates the relative contributions from Ni atoms in the outer and inner layers.}
\label{fig1}
\end{figure}

To investigate the band structure of La$_4$Ni$_3$O$_{10}$ under ambient pressure, we first perform DFT calculations using the projector-augmented wave (PAW) pseudo-potentials with the exchange–correlation of the Perdew–Burke–Ernzerhof (PBE) and the GGA+U approach, as implemented in the Vienna ab-initio Simulation Package (VASP)~\cite{dft1,dft2,dft3,dft4}. The Hubbard $U$ is set to be 3.5 eV to account for the correlation effects of Ni-3d electrons~\cite{yang2023arpes}. We adopt the $P2_1/a$ crystal structure with lattice constants measured experimentally at 0 GPa ~\cite{li2024structural}, and relax the atomic positions until the atomic force on each atom is less than $10^{-3}$ eV/$\dot{\text{A}}$. As shown in Fig. 1(a) and (b), the low-energy electronic states mainly originate from the Ni-$3d_{z^2}$ and $3d_{x^2-y^2}$ orbitals, which allows us to construct high-quality, maximally localized Wannier representations~\cite{mostofi2008wannier90} by projecting the Bloch states (with a 14$\times$14$\times$6 $k$ mesh) from the DFT calculations onto the Ni-$3d_{z^2}$ and $3d_{x^2-y^2}$ orbitals.

The obtained TB Hamiltonian in real space can be expressed as
\begin{align}
H_{{\rm TB}}
=\sum_{\bm r_i\Delta \bm r\mu\nu\sigma}t_{\Delta \bm r\mu\nu}
c^{\dagger}_{\bm r_i\mu\sigma}c_{(\bm r_i+\Delta \bm r)\nu\sigma},
\end{align}
Here $\bm r_i$ represents the coordinates of site $i$, $\Delta \bm r_{x}(\Delta \bm r_{y})\in(-3,3)$ represents the hopping range. The indices $\mu, \nu=1, \cdots, 12$ containing the $d_{z^2}$ and $d_{x^2-y^2}$ orbitals of the top-, middle- and bottom-layer Ni$^\mathrm{A}$ (Ni$^\mathrm{B}$), and $\sigma =\uparrow, \downarrow$ labels spin. $c^{\dagger}_{\bm r_i\mu\sigma}(c_{\bm r_j\nu\sigma})$ creates (annihilates) a spin-$\sigma$ electron in the orbital $\mu(\nu)$ at site $i(j)$. The elements of the hopping matrix $t_{\Delta \bm r\mu\nu}$ of this twelve-band tight-binding model are provided in the SM~\cite{SM}.

The obtained band structure for this TB model is shown in Fig. \ref{fig1}(c). A comparison between this band structure and the DFT one, shown in Fig. \ref{fig1}(a), indicates that
the essential feature of the DFT results are well captured. The corresponding FS is shown in Fig. \ref{fig1}(d). The FS features two electron pockets, $\alpha_1$ and $\alpha_2$, as well as two hole pockets, $\beta_1$ and $\beta_2$. Both the $\alpha_1$ and $\beta_1$ pockets primarily originate from the outer layer orbitals, with the $\alpha_1$ pocket being entirely formed by the outer 
$d_{z^2}$ orbital. The $\beta_1$ pocket consists of contributions from both the $d_{z^2}$ and $d_{x^2-y^2}$ orbitals. In contrast, the primary orbital components of the 
$\alpha_2$ and $\beta_2$ pockets are predominantly derived from the inner layer $d_{x^2-y^2}$ orbitals. The overall electronic structure is in good agreement with 
previous ARPES experiment\cite{zhang2023superconductivity,li2017fermiology,du2024correlated}.

\section{SDW Calculation Using RPA}
\label{sec:SDW}
\begin{figure}[htbp]
\centering
\includegraphics[width=0.45\textwidth]{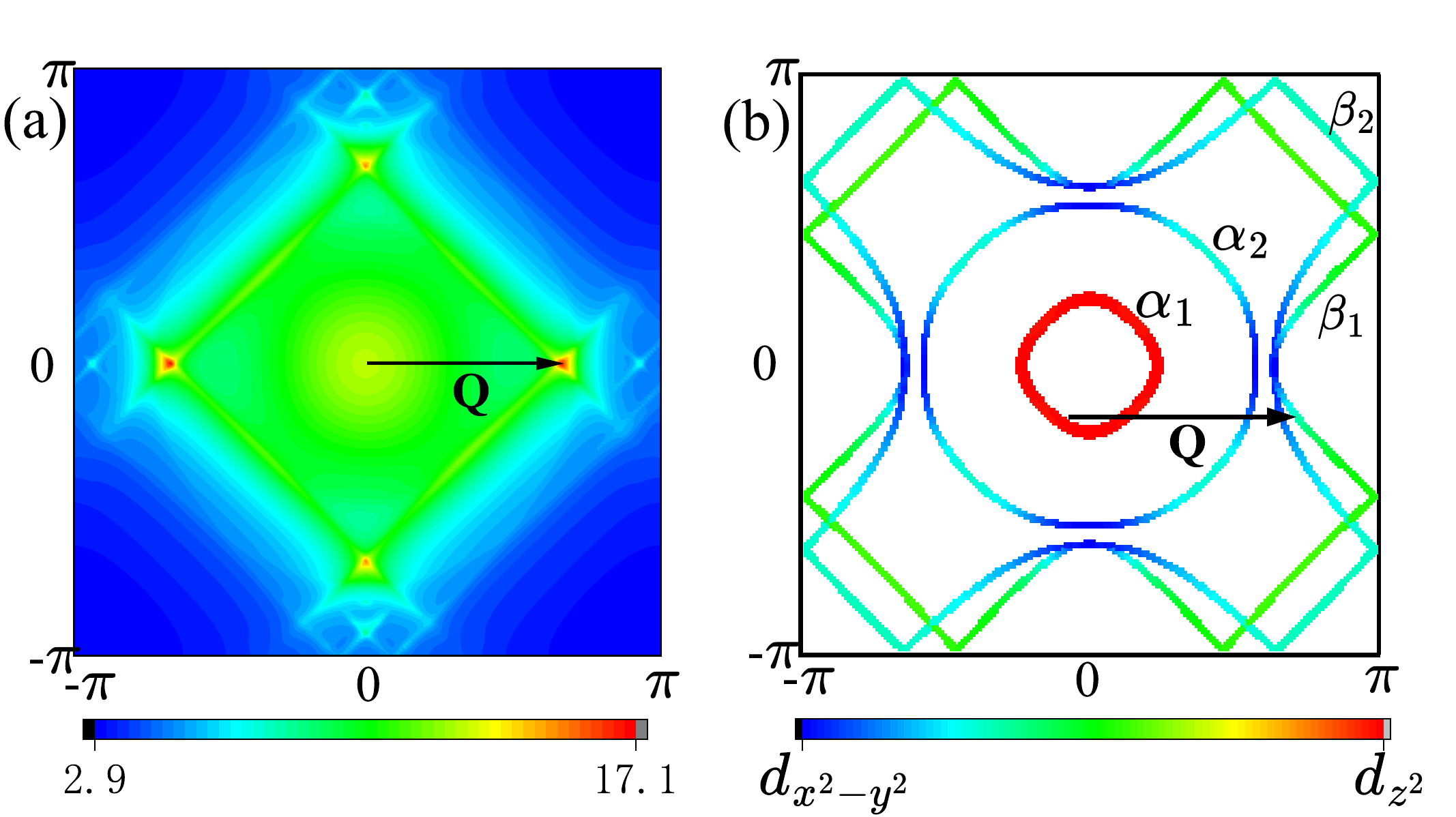}
\caption{(color online) (a) The distribution of the RPA-renormalized spin susceptibility $\chi^{(s)}(\mathbf{q})$ in the BZ for $U=1.3$ eV and $J_H$=$0.25U$. The maximum of the distribution is located at $\mathbf{Q}\approx(\pm 0.7\pi,0)$. (b) FS nesting connected by the $\mathbf{Q}$ vector. The color scheme in (b) indicates the relative contributions of the $d_{x^2-y^2}$ and $d_{z^2}$ orbitals. }
\label{fig2}
\end{figure}

\begin{figure*}[htbp]
\centering
\includegraphics[width=0.75\textwidth]{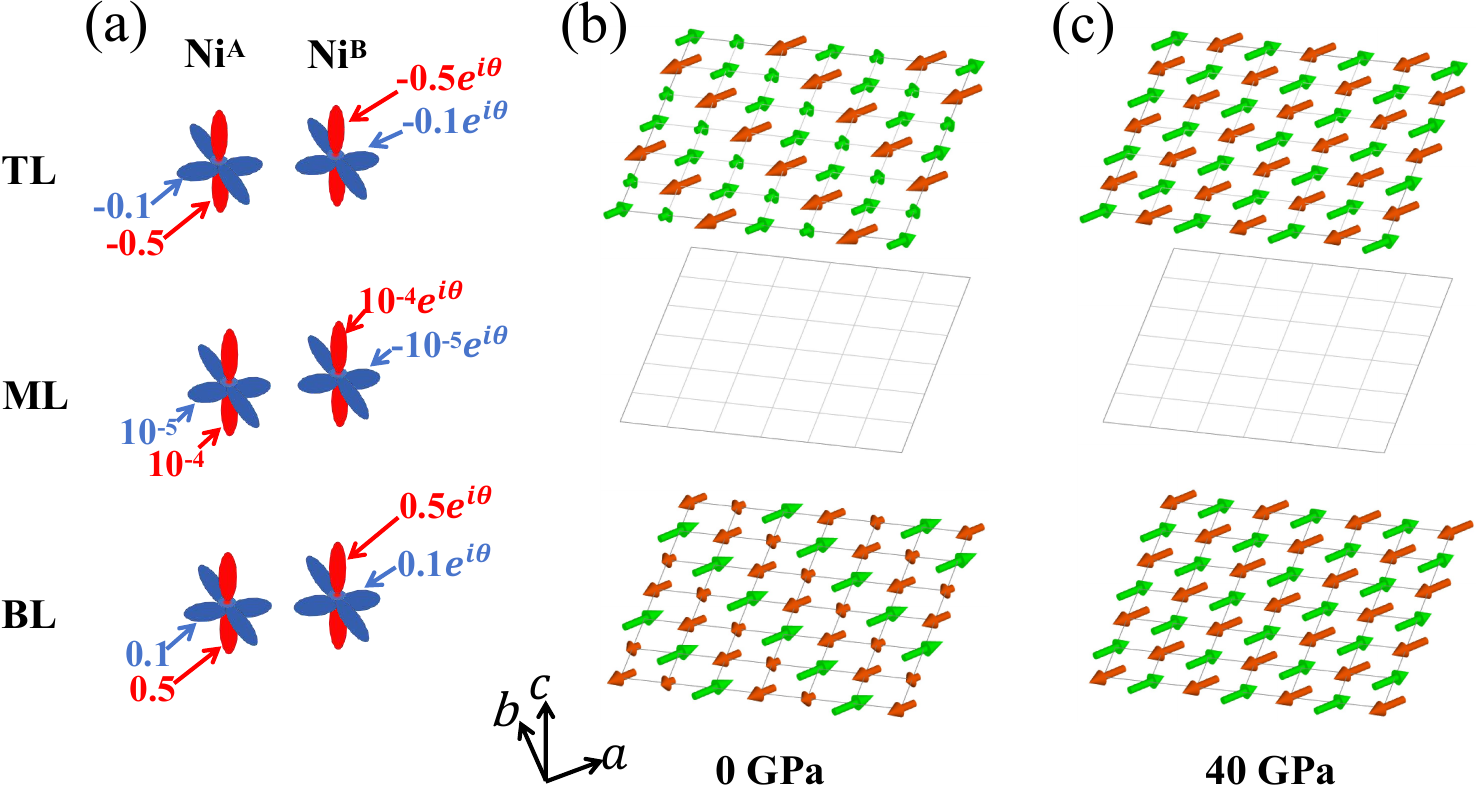}
\caption{(color online) (a) The leading spin-fluctuation pattern within an unit cell, where the red (blue) pattern represents the Ni$-d_{z^2}(d_{x^2-y^2})$ orbital. The trilayer is labeled as TL, ML, and BL. The numbers indicate the magnetic moment strength of the corresponding orbitals, and the phase angle of the Ni$^{\mathrm{B}}$ sublattice is $\theta=0.7\pi$. (b)
SDW model at 0 GPa plotted in the real space. (c) SDW model at 40 GPa plotted in the real space. Arrows represent the magnetic moments, with green and red indicating opposite directions. The magnitude of each magnetic moment is reflected by the arrow length. The absence of arrows in the middle layer indicates a very small magnetic moment. With SU(2) symmetry, the magnetic moments may point along the $x$, $y$, or $z$-axis.}
\label{fig3}
\end{figure*}

\begin{figure}[htbp]
\centering
\includegraphics[width=0.45\textwidth]{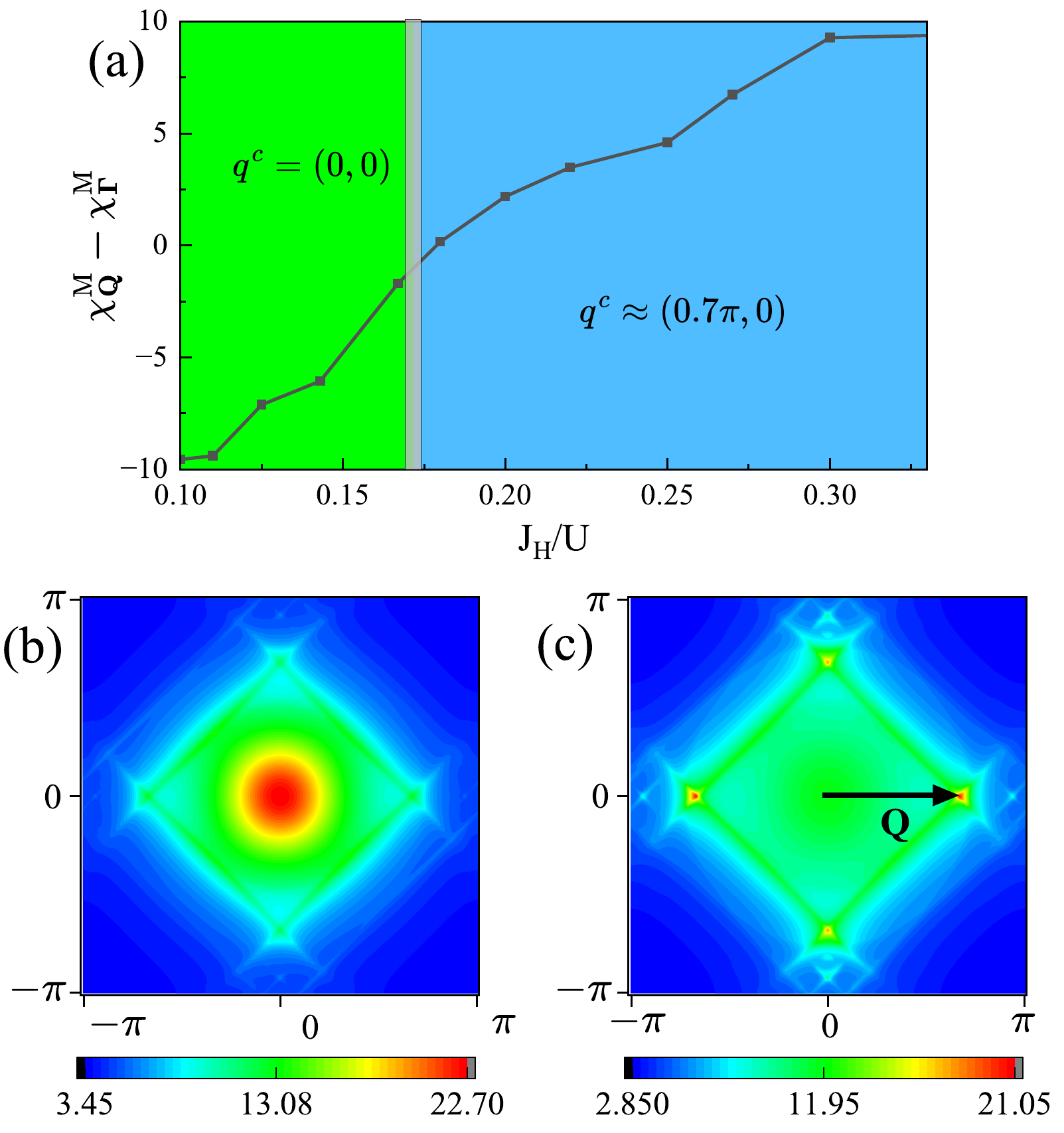}
\caption{(color online) (a) Dependence of $\chi^{M}_{\mathbf{Q}}-\chi^{M}_{\Gamma}$ on $J_H$ with $U\approx U_c$. Here, $\chi^{M}_{\mathbf{Q}}$ and $\chi^{M}_{\Gamma}$ represent the maximum eigenvalues of $\chi^s$ at the $\mathbf{Q}$-point $\Gamma$-point, respectively. $\bm{q}^c$ denotes the momentum corresponding to the strongest spin susceptibility in the first BZ. When $\chi^{M}_{\mathbf{Q}}-\chi^{M}_{\Gamma}<0$, $\bm{q}^c$ is located at the $\Gamma$-point; when $\chi^{M}_{\mathbf{Q}}-\chi^{M}_{\Gamma}>0$, $\bm{q}^c$ is located at the $\mathbf{Q}$-point. (b) Distribution of $\chi^{(s)}(\mathbf{q})$ in the BZ for $J_H = 0.1U$. (c) Distribution of $\chi^{(s)}(\mathbf{q})$ in the BZ for $J_H = 0.3U$, both with $U \approx U_c$. }
\label{fig4}
\end{figure} 

\begin{figure}[htbp]
\centering
\includegraphics[width=0.44\textwidth]{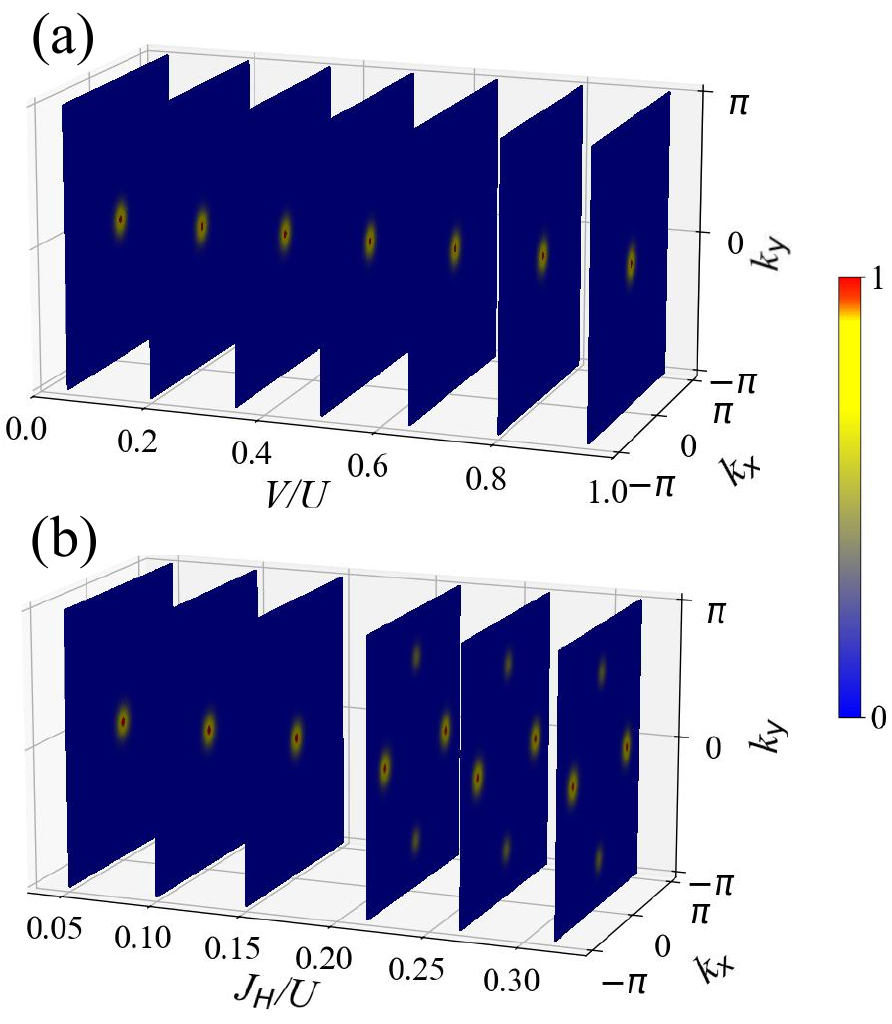}
\caption{(color online) Distribution of $\chi^s(\bm k)$ in the BZ under different conditions. (a) Variation of $\chi^s(\bm k)$ with $V$ when $J_H=0$, (b) Variation of $\chi^s(\bm k)$ with $J_H$ when $V=0$. Bright spots indicate the location of $\bm{q}^c$, and $U\approx U_c$ ensures the system is near the critical interaction strength. }
\label{fig5}
\end{figure}

To investigate the SDW formed by electron interactions, we consider the repulsive Hubbard model described by Eq. \eqref{hubbard}:
\begin{align}\label{hubbard}
H_{int}&=U\sum_{i\mu}n_{i\mu\uparrow}n_{i\mu\downarrow}+
V\sum_{i,\sigma,\sigma^{\prime}}n_{i1\sigma}n_{i2\sigma^{\prime}}+J_{H}\sum_{i\sigma\sigma^{\prime}} \nonumber\\
&\Big[c^{\dagger}_{i1\sigma}c^{\dagger}_{i2\sigma^{\prime}}c_{i1\sigma^{\prime}}c_{i2\sigma}+(c^{\dagger}_{i1\uparrow}c^{\dagger}_{i1\downarrow}c_{i2\downarrow}c_{i2\uparrow}+h.c.)\Big]
\end{align}
Here, $U$, $V$, and $J_H$ denote the intra-orbital, inter-orbital Hubbard repulsion, and the Hund's rule coupling (and the pair hopping) respectively, which satisfy the relation $U=V+2J_H$. We employ the multi-orbital RPA approach \cite{takimoto2004strong,yada2005origin,kubo2007pairing,graser2009near,liu2013d+,zhang2022lifshitz,kuroki101unconventional} to treat this Hamiltonian. By renormalization, the spin susceptibility $\bm{\chi}^{(s)}$ and charge susceptibility $\bm{\chi}^{(c)}$ can be defined as Eq. \eqref{chisce}:
\begin{align}\label{chisce}
 \bm{\chi}^{(s)}(\bm {k},i\nu)=[I-\bm{\chi}^{(0)}(\bm {k},i\nu)
 U^{(s)}]^{-1}\bm{\chi}^{(0)}(\bm {k},i\nu),\nonumber\\
 \bm{\chi}^{(c)}(\bm {k},i\nu)=[I+\bm{\chi}^{(0)}(\bm {k},i\nu)
 U^{(c)}]^{-1}\bm{\chi}^{(0)}(\bm {k},i\nu).
\end{align}
 Where $\bm{\chi}^{(0)}$ is bare susceptibility for the non-interacting case, expressed as a tensor $\bm{\chi}^{(0)pq}_{st}$, with $p,q,s,t$ as orbital indices. Similarly, $\bm{\chi}^{(s)}$ can be expressed as $\bm{\chi}^{(s)pq}_{st}$. $U^{(s,c)}$ is the renormalized interaction strength, represented as a $12^2\times12^2$ matrix in this twelve-orbital system. Note that there is a critical interaction strength $U_c^{(s,c)}$ for both spin and charge susceptibilities. When $U\geq U_c^{(s,c)}$, the denominator matrix in Eq. \eqref{chisce} will have zero eigenvalues for certain values of $\bm{k}$, causing the renormalized spin or charge susceptibility to diverge. This divergence indicates the onset of magnetic or charge order. When fixing $J_H = 0.25U$, we find that $U_c^{(s)}\approx 1.33$ eV. Defining $\chi^s(\bm {k})$ as the maximum eigenvalue of $\bm{\chi}^{(s)}$ at each momentum, Fig. \ref{fig2}(a) shows its distribution in the BZ for $U=1.3$ eV$<U_c^{(s)}$. Notably, the strongest spin susceptibility is located near the momentum $\mathbf Q\approx(\pm 0.7\pi,0)$, which is very close to the experimentally measured value of $(\pm 0.76\pi,0)$\cite{zhang2020intertwined}. At ambient conditions, La$_4$Ni$_3$O$_{10}$ crystallizes in a monoclinic structure with the space group $P_{21/a}$, which weakly breaks the $D_4$ symmetry and restricts the wave vector $\mathbf Q$ to just two points in the first BZ. In the FS, $\mathbf Q$ is the nesting vector between the $\alpha_1$-pocket and $\beta_1$-pocket, as shown in Fig. \ref{fig2}(b).
 

The spatial distribution of the dominant spin fluctuations' intra-unit-cell modulation is illustrated in Fig. \ref{fig3}(a). This distribution is determined by the eigenvector corresponding to the largest eigenvalue of the spin susceptibility matrix defined as $\bm \chi^{(s)}_{p,s}\equiv \bm \chi^{(s)pp}_{ss}(\bm k)$ with $\bm k=(\pm 0.7\pi,0)$. As shown in Fig. \ref{fig3}(a), the two outer planes are out-of-phase, with the magnetic moments of Ni$^{\mathrm{A}}$ and Ni$^{\mathrm{B}}$ exhibiting a phase difference of 0.7$\pi$. The magnetic moment in the middle layer (ML) is nearly zero, indicating that the SDW has a node at the ML. In the top layer (TL) and bottom layer (BL), the magnetic moments of the $d_{z^2}$ orbitals are significantly larger than those of the $d_{x^2-y^2}$ orbitals, which are approximately one-fifth as strong. This observation suggests that spin fluctuations primarily occur in the $d_{z^2}$ orbitals of the two outer layers, consistent with previous DMFT calculation results\cite{huo2024electronic,Leonov20244310,tian2024effective,wang2024non}.

The spin-fluctuation pattern of inter-unit-cell is determined by $\mathbf{Q}$. The main features of the SDW in real space can be captured by $\phi(\bm r)=\sum_{\bm k}C_z|e^{i\bm k \cdot \bm r}|\cos(\mathbf{Q}\cdot \bm{r})$, where $C_z=1,0,-1$ representing the magnetic stacking pattern along the $c$-axis, and the summation over $\bm k$ is normalized by their number. The SDW has a node on the inner plane and is antiphase between the two outer planes, as shown in Fig. \ref{fig3}(b), which is consistent with experimentally observed features\cite{zhang2020intertwined}. Additionally, the SDW period of 1.5$\mathbf a$ closely matches the experimental value\cite{zhang2020intertwined}. This contrasts with the Néel-type SDW shown in Fig. \ref{fig3}(c) for high-pressure electron-doped La$_4$Ni$_3$O$_{10}$ \cite{zhang2024s}.

When adjusting $J_H$, we find that the momentum of the strongest spin susceptibility, $\bm{q}^c$, transitions from $\mathbf{Q}\approx(\pm 0.7\pi,0)$ to $\Gamma=(0,0)$ as shown in Fig. \ref{fig4}(a). This transition is further illustrated in the distribution of $\chi^s(\mathbf{q})$ in the BZ. For $J_H = 0.1U$, as shown in Fig. \ref{fig4}(b), $\bm{q}^c$ is located at the $\Gamma$-point. However, when $J_H = 0.3U$, $\bm{q}^c$ shifts to the $\mathbf{Q}$-point, as shown in Fig. \ref{fig4}(c). To further investigate this transition, we calculated the dependence of $\bm{q}^c$ on $V$ for $J_H$=0 (Fig. \ref{fig5}(a)) and the dependence of $\bm{q}^c$ on $J_H$ for $V$=0 (Fig. \ref{fig5}(b)), excluding the commonly used relation $U = V + 2J_H$ to pinpoint the key parameter. We observe that $\bm{q}^c$ remains located at $\Gamma$ within the range $V\in(0,U)$, but shifts from $\Gamma$ to $\mathbf{Q}$ for $J_H>0.16U$. 

Regarding the distribution of spin fluctuations in the intra-unit-cell at $\bm{q}^c=(0,0)$, our calculations reveal that the magnetic moments are predominantly localized in the two outer $d_{z^2}$ orbitals and are antiphase between the two outer planes. The phase difference between Ni$^{\mathrm{A}}$ and Ni$^{\mathrm{B}}$ in this case is $\pi$. Notably, this configuration exhibits the same real-space SDW features as those shown in Fig. \ref{fig3}(c) for the high-pressure case. This indicates that for small $J_H$, the magnetic fluctuation resembles the Néel-type order observed under high pressure. In contrast, for $J_H > 0.16U$, the system transitions to the stripe SDW order characterized by $\mathbf{Q} \approx (\pm 0.7\pi, 0)$ .

\section{Comparison with High Pressure}
\label{sec:CompHP}
\begin{figure}[htbp]
\centering
\includegraphics[width=0.45\textwidth]{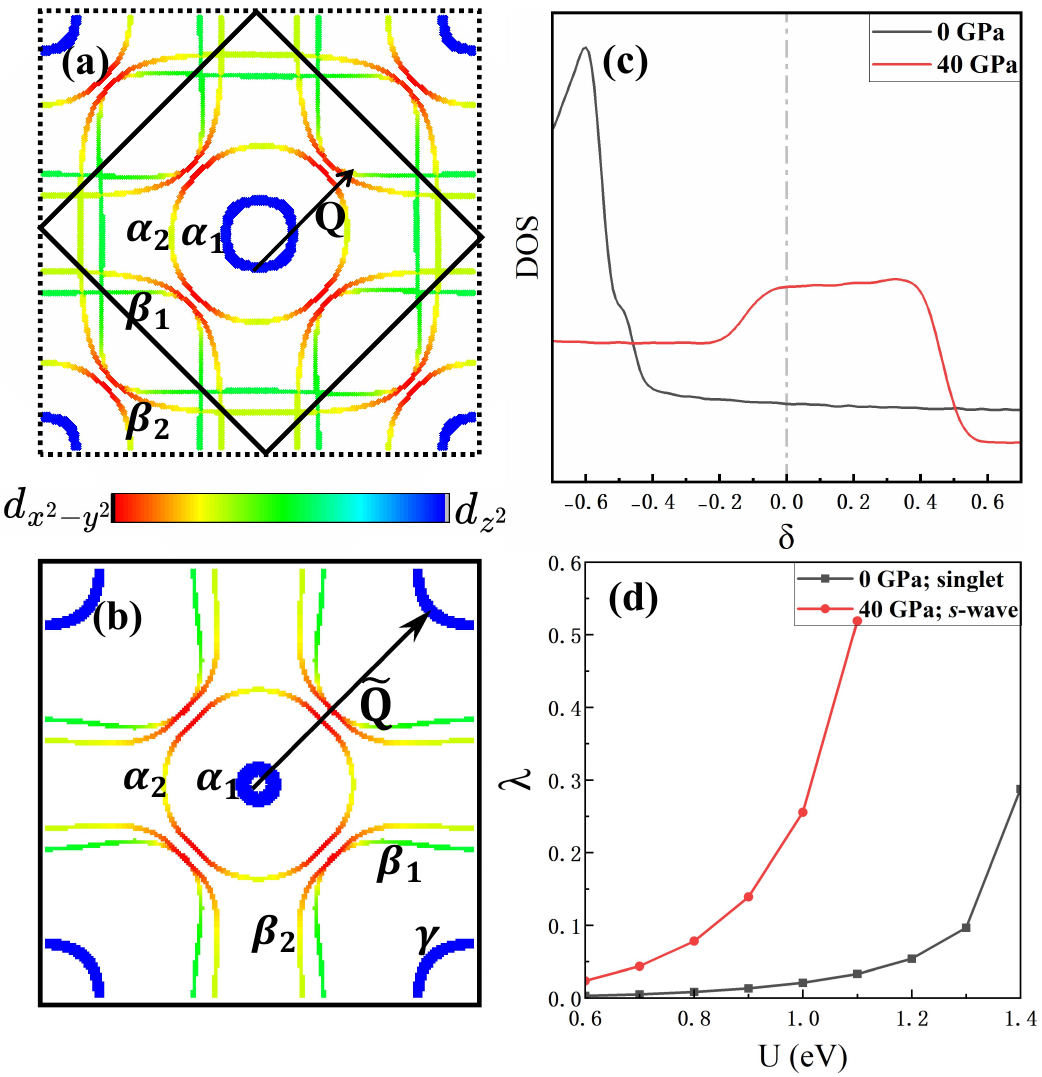}
\caption{(color online) Comparison between ambient and high pressure. (a) and (b) show the FS at 0 GPa and 40 GPa, respectively. The solid black line outlines the first BZ, while the area between the solid and dashed lines in (a) represents the second BZ. The color scheme in (a) and (b) indicate the relative contributions of the $d_{x^2-y^2}$ and $d_{z^2}$ orbitals. (c) DOS, with the horizontal axis representing doping levels; the gray dashed line corresponds to the undoped case. (d) The largest pairing eigenvalue, $\lambda$, as a function of $U$, with $J_H=U/6$ fixed for undoped La$_4$Ni$_3$O$_{10}$. In (c) and (d), the black solid line represents the 0 GPa case, while the red solid line denotes the 40 GPa case.}
\label{fig6}
\end{figure}

There are two nesting vectors that the system tends to form: one is $\mathbf{Q}\approx(\pm 0.7\pi,0)$ between the $\alpha_1$ pocket and the $\beta_1$ pocket, while the other is near (0,0). Notably, the band structure at the $\Gamma$ point exhibits two nearly touching regions: the flat band bottom and flat band top, both near the Fermi level, as shown in Fig. \ref{fig1}(c). The flat band top forms the $\gamma$ pocket at high pressure, with its flat segment located at the M point in the unfolded BZ. The nesting between the $\gamma$ pocket and the $\alpha_1$ pocket is $(\pi,\pi)$\cite{zhang2024s,zhang2024prediction}. After BZ folding, the flat band segment shifts from the M point to the $\Gamma$ point, meaning the $(\pi,\pi)$ nesting vector at high pressure corresponds to the current (0,0) nesting vector. The dominant nesting vector varies with $J_H$. When $J_H<0.16U$, the nesting vector is (0,0), while it shifts to ($\pm$0.7,0) for $0.16U<J_H<0.33U$. Based on experimental observations, at ambient pressure, the ($\pm$0.7,0) nesting in La$_4$Ni$_3$O$_{10}$ predominates over the (0,0) nesting. 

To compare with the characteristics under high pressure, we investigated the FS nesting, DOS, and superconducting pairing strength. The FS at 0 GPa and 40 GPa are shown in Fig. \ref{fig6}(a) and (b), respectively. Due to the expansion of the unit cell at 0 GPa, which is $\sqrt{2}\times\sqrt{2}$ times larger than at 40 GPa, the first BZ is reduced in size. Consequently, the second BZ is also shown in (a) to clearly illustrate the folded $\alpha_1$, $\alpha_2$, $\beta_1$ and $\beta_2$ pockets. The vectors $\mathbf{Q}$ and $\mathbf{\tilde{Q}}$ represent the dominant nesting wave vectors at 0 GPa and 40 GPa, respectively. The $\mathbf{Q}$ vector connects the $\alpha_1$ pocket and the $\beta_1$ pocket, with the $\alpha_1$ pocket formed by the outer-layer $d_{z^2}$ orbital. Although the $\beta_1$ pocket includes a mixture of the outer-layer $d_{x^2-y^2}$ and $d_{z^2}$ orbitals, the nesting with the $\alpha_1$ pocket is primarily driven by the outer-layer $d_{z^2}$ orbital. For the $\mathbf{\tilde{Q}}$ vector, high-pressure calculations show that both the $\alpha_1$ pocket and the $\gamma$ pocket are associated with the outer-layer $d_{z^2}$ orbital. Therefore, whether under high pressure or ambient pressure, the nesting vector is formed between the pockets associated with the outer-layer $d_{z^2}$ orbitals. These nesting contribute to the SDW pattern, leading to a stripe-type SDW at ambient pressure, as shown in Fig. \ref{fig3}(b), and a Néel-type SDW at high pressure, as shown in Fig. \ref{fig3}(c).

Since the DOS at 0 GPa is lower than that at 40 GPa when doping is zero, as show in Fig. \ref{fig6}(c), it is not surprising that superconductivity was initially discovered under high-pressure conditions. After performing the RPA, we used the mean-field approximation to calculate the linearized superconducting gap equation. The resulting dependence of the superconducting pairing eigenvalue $\lambda$ on $U$ is shown in Fig. \ref{fig6}(d). Here $\lambda$ reflects the level of superconducting $T_c$, where a larger $\lambda$ implies a higher $T_c$ (for more details, see the next section). Compared to 40 GPa, the $\lambda$ at 0 GPa is much lower, indicating that it is difficult to experimentally detect superconductivity in undoped La$_4$Ni$_3$O$_{10}$ at ambient pressure.

\section{Superconductivity Induced by Doping}
\label{Doping}
\begin{figure}[htbp]
\centering
\includegraphics[width=0.45\textwidth]{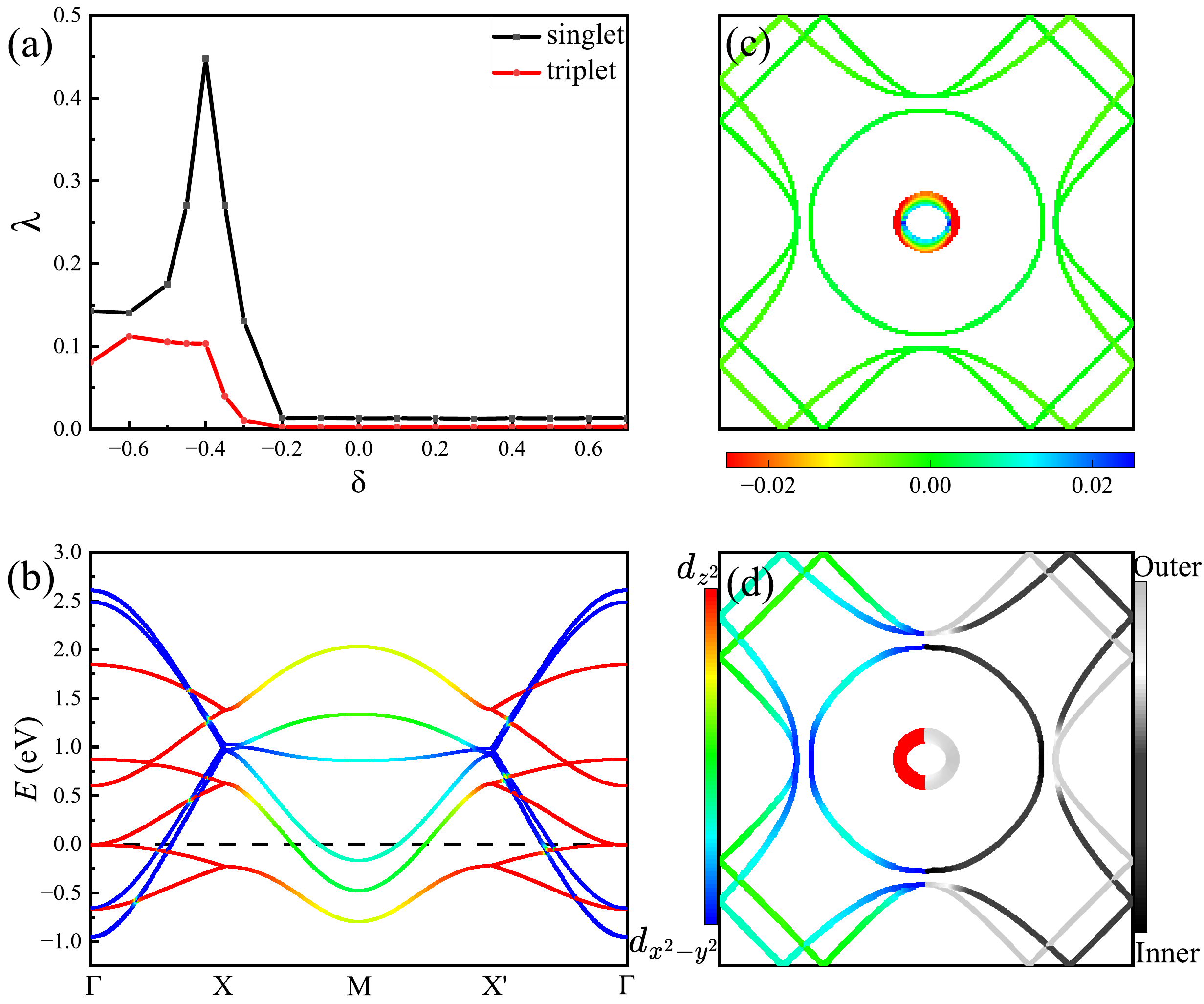}
\caption{(color online) Doping effects at ambient pressure. (a) The largest pairing eigenvalues $\lambda$ as a function of doping, with fixed $J_H=U/6$ for La$_4$Ni$_3$O$_{10}$ at 0 GPa. The black solid line represents the singlet pairing state, while the red solid line represents the triplet pairing state. (b) Band structure of $\delta=-0.4$. (c) Distribution of the dominant gap at the FS for $\delta=-0.4$. (d) FS of $\delta=-0.4$. }
\label{fig7}
\end{figure}

To investigate whether ambient pressure can achieve an appropriate superconducting $T_c$, we consider the doped case. In the undoped scenario, the number of electrons per unit cell is $n=8$. We vary the electron count to $n\in(7.3,8.7)$, corresponding to $\delta\in(-0.7,0.7)$. The doping-dependent of DOS is shown in Fig. \ref{fig6}(c). In the range of (-0.4, 0.4), the DOS at ambient pressure is consistently lower than that at high pressure. Upon further hole doping at $\delta$=-0.4, the DOS at ambient pressure begins to rise significantly, eventually surpassing the high-pressure DOS and reaching a peak at $\delta$=-0.6.

After applying the RPA, we rewrite the interaction in Eq. \eqref{hubbard} as an effective interaction $V(\bm k,\bm q)$. While the mean-field method could be used directly to solve the linearized gap equation for $V(\bm k,\bm q)$ and determine the leading pairing symmetry and superconducting $T_c$, we choose to use a thin-shell approach for computational convenience. This leads to the following eigenvalue problem of $V(\bm k,\bm q)$ as:
\begin{equation}\label{gap}
-\frac{1}{(2\pi)^2}\sum_{\beta}\oint_{FS}
dq_{\Vert}\frac{V^{\alpha\beta}(\bm {k},\bm{q})}{v^{\beta}_{F}(\bm{q})}\Delta_{\beta}(\bm{q})=\lambda
\Delta_{\alpha}(\bm {k})
\end{equation}
Here, $\Delta_{\alpha}(\bm {k})$ represents the relative gap function on the $\alpha$-th FS patch near $T_c$, and the eigenvalue $\lambda$ is related to $T_c$ through $\lambda^{-1}=\ln \left(1.13 \frac{\hbar \omega_{D}}{k_{B}} T_{c}\right)$. The leading pairing symmetry is determined by the largest eigenvalue $\lambda$. To facilitate an effective comparison with the high-pressure scenario, we set $U=1$ eV and $J_H=U/6$, following Ref. \cite{zhang2024s}, and then calculated the dependence of $\lambda$ on doping $\delta$, as shown in Fig. \ref{fig7}(a). It is evident that singlet pairing is always favored over triplet pairing. When $\delta>-0.2$, the system exhibits a measurable $\lambda$, and at $\delta=-0.4$, $\lambda$ reaches its maximum of approximately 0.45. Combining this with the high-pressure results in Fig. \ref{fig6}(d) indicates that the $\lambda$ at this doping is comparable to that under high pressure, suggesting that doping at $\delta=-0.4$ can achieve superconductivity at ambient pressure.

At $\delta=-0.4$, the Fermi level drops to the nearly touching point of the flat band bottom and top near the $\Gamma$ point, as shown in Fig. \ref{fig7}(b). This configuration likely optimizes the nesting around the $\Gamma$ point, with the nesting vector being (0,0), allowing $\lambda$ to reach its peak at this doping level. Interestingly, under high pressure, this nesting vector transforms into $(\pi,\pi)$, connecting the $\alpha_1$ pocket and the $\gamma$ pocket, as shown in Fig. \ref{fig6}(b). Current theoretical calculations under high pressure suggest that the superconducting gap is primarily distributed in these two pockets\cite{zhang2024s,zhang2024prediction}. Additionally, we examined cases with larger $J_H$. For $J_H=0.25U$ and $J_H=0.3U$, the nesting vector remained fixed at (0,0), indicating that the nesting between the $\alpha_1$ pocket and the $\gamma$ pocket remains robust despite the increase in $J_H$.

The distribution of the relative gap function on the FS is shown in Fig. \ref{fig7}(c). Under high pressure, the system exhibits $D_4$ symmetry, which permits potential pairing symmetries such as $s$-wave, $d_{xy}$-wave, $d_{x^2-y^2}$-wave and degenerate $(p_x,p_y)$-wave states. At ambient pressure, however, the $D_4$ symmetry is slightly broken, reducing it to $D_2$ symmetry. As a result, the possible superconducting pairing symmetries are limited to singlet and triplet states. The gap shown in Fig. \ref{fig7}(c) exhibits singlet symmetry, consistent with the dominance of the singlet $\lambda$ observed in Fig. \ref{fig7}(a).
The superconducting gap is primarily concentrated in $\alpha_1$ pocket. As shown in Fig. \ref{fig7}(d), this pocket is composed of pure outer-layer $d_{z^2}$ orbitals, consistent with the reported characteristics of the superconducting gap distribution in high-pressure La$_4$Ni$_3$O$_{10}$ \cite{zhang2024s,zhang2024prediction}.

\section{Conclusion}
\label{sec:conclusion}
Adopting the TB model fitted from the DFT band structure, we investigated the SDW and potential superconductivity of La$_4$Ni$_3$O$_{10}$ at ambient pressure using the RPA approach. We obtain a SDW pattern consistent with experimental observations, primarily driven by Hund’s coupling. This conclusion is supported by comparisons with results obtained under reduced Hund’s coupling. When $J_H>0.16$, the system exhibits a stripe SDW pattern, characterized by in-plane antiferromagnetism with a wave vector $(\pm 0.7\pi, 0)$ and antiphase antiferromagnetic ordering between the two outer layers. 

This stripe SDW arises primarily from the magnetic moments in the two outer $d_{z^2}$ orbitals, with the top and bottom layers adopting an interlayer antiferromagnetic configurations, while the middle layer serves as an SDW node. The wave vector $(\pm 0.7\pi,0)$ corresponds to the nesting vector connecting the $\alpha_1$ pocket and the $\beta_1$ pocket. In the undoped case, the (0,0) nesting vector—equivalent to $(\pi,\pi)$ in the unfolded high-pressure BZ—is suppressed by $J_H$.

With $\delta=-0.4$ hole doping, the system transitions into a superconducting phase. Notably, both ambient- and high-pressure superconductivity share a key characteristic: the two pockets connected by nesting are formed by pure outer-layer $d_{z^2}$ orbitals. Furthermore, this nesting is remarkably stable and remains unaffected by $J_H$. Since the eigenvalue $\lambda\approx0.45$ at this doping exceeds the high-pressure value of $\lambda=0.25$, we anticipate that hole doping will induce superconductivity in La$_4$Ni$_3$O$_{10}$ at ambient pressure, with $T_c$ at $\delta=-0.4$ expected to surpass that at high pressure.

\begin{acknowledgements}
We are grateful to the discussions with Chen Lu. This work is supported by the NSFC under Grant Nos.12234016, 12074031, 12141402, and 12334002, Guangdong province (2020KCXTD001), Shenzhen Science and Technology Program under Grant No. RCJC20221008092722009. W.-Q. Chen is supported by the National Key R\&D Program of China (Grants No. 2022YFA1403700), the Science, Technology and Innovation Commission of Shenzhen Municipality (No. ZDSYS20190902092905285), and Center for Computational Science and Engineering of Southern University of Science and Technology.
\end{acknowledgements}

\end{document}